\documentstyle[twoside,fleqn,espcrc2]{article}

\newcommand{\AmS}{{\protect\the\textfont2
  A\kern-.1667em\lower.5ex\hbox{M}\kern-.125emS}}

\title{
\begin{flushright}
\begin{small}
SU-ITP-97-39,~hep-th/9709202\\
\end{small}
\end{flushright}
\vspace{1.cm}
Quantization of p-branes, D-p-branes and M-branes \thanks{To appear in the Proceedings of Strings 97, Amsterdam, June 1997. This work is supported by the NSF  grant PHY-9219345 and a NATO Collaboration Research Grant. }
}

\author{R. Kallosh\\
Physics Department, Stanford University, Stanford, CA
94305-4060
}%

\begin{document}

\begin{abstract}
Killing spinors of space-time BPS  configurations play an important role in quantization of theories with the fermionic worldvolume local symmetry. We show  here how it works for the GS superstring,  BST
supermembrane and 
M-5-brane.  We show that the non-linear generalization of the (2,0) d=6 tensor supermultiplet action  is the  M-5-brane action in a Killing gauge. For D-p-branes the novel feature of quantization is  that they can be quantized  Lorentz covariantly, in particular,  for D-0-brane a gauge exists where the action is  covariant and free. We present a general condition on possible choice of gauges for the $\kappa$-symmetric branes.

\end{abstract}

\maketitle

We review here some work on quantization of the local fermionic worldvolume symmetry presented in \cite{duality,bk,kallosh,KalM5}.
This is a generalized version of the talk presented at Strings 97 in Amsterdam in June 1997.  We give a short summary of the new results in the end of this paper.

\section{WORLDVOLUME ACTIONS AND THEIR LOCAL SYMMETRIES}

The actions of Green-Schwarz string in D=10  target space and of the Bergshoeff-Sezgin-Townsend
membrane in D=11 target space have local fermionic $\kappa$-symmetry, reparametrization invariance and manifest  space-time supersymmetry. They depend on worldvolume fields $Z^M=( X^m (\xi) ,  \theta^\alpha (\xi) )$ related to coordinates of the space-time superspace. Here $X^m, m =0,1, \dots, D-1$ are bosonic and $\theta^\mu , \mu = 1, \dots, 32$ are fermionic coordinates of the superspace. The bosonic coordinates on the brane are $\xi^i , i=0, \dots , p$. For the string $p=1$ and for the membrane $p=2$.
Both actions belong to a general class of $\kappa$-symmetric p-brane actions and upon gauge-fixing describe the scalar supermultiplets of worldvolume supersymmetry. 

D-p-brane actions \cite{Ce1,Ag1,Be1} depend in addition on world-volume  1-form gauge potential $A_{i}$ and upon gauge-fixing describe the vector multiplets of worldvolume supersymmetry. 

Finally, the M-5-brane action \cite{5b,ag}
depends on a 2-form gauge potential $A_{ij}$ with the self-dual field strength. This action after gauge-fixing describes the tensor multiplets of worldvolume supersymmetry.

The class of actions we consider  have 32-dimensional
global space-time supersymmetry   and the local fermionic $\kappa$-supersymmetry :
\begin{eqnarray}
&&\delta_\epsilon \theta = \epsilon, \qquad \delta_\epsilon X^m = \bar\epsilon
\Gamma^m
\theta \ , \label{susytrans}\\
\nonumber\\
&&\delta_{\kappa} \theta= (1+\Gamma)\,   \kappa (\xi)\ , \nonumber\\
\nonumber\\
&& \delta_{\kappa}  X^m = \bar\theta \Gamma^m \delta\theta_{\kappa} \label{Xtrans}  \ , \dots \ .
\end{eqnarray}
Both $\epsilon$ and $\kappa$ are 32-dimensional, but $\epsilon$ is global and $\kappa$ depends on coordinates of the brane $\xi$.
Here dots mean the transformations of Born-Infeld or a tensor field. $ \Gamma$
is a function of the fields of the brane 
$
\Gamma = \Gamma \left (Z(\xi), A(\xi)\right)
$
and depends on $\xi^i$ therefore. The
matrix  $\Gamma(\xi)$  squares to 1 and has a vanishing trace:
therefore $(1+ \Gamma) (1- \Gamma)=0
$, 
i.e.  $1+ \Gamma$ is a projector which makes a 32-dimensional parameter of
$\kappa$-supersymmetry effectively only 16-dimensional. This allows to exclude half of the original $\theta$ variables in the classical action from the quantized action. 

Local reparametrization symmetry of the worldvolume includes transformations of the type $\delta Z^M (\xi)= \delta \xi^i \partial_i Z^M (\xi)$ for the scalars on the worldvolume, etc. Here $\delta \xi^i(\xi)$ depends on coordinates of the worldvolume, i.e. it is a gauge symmetry, which allows us to exclude some bosonic fields from the classical action.

Quantization, or gauge-fixing of local symmetries of the worldvolume actions will be described  here in detail in a Killing adapted gauge but also in most general gauges \cite{KalM5}. The local symmetries will be used to exclude the unphysical degrees of freedom. The resulting worldvolume actions have 
some particular form of the global supersymmetry, which is a combination of the original space-time supersymmetry and a special choice of $\kappa$-symmetry which preserves the Killing spinor adapted gauge. The advantage of using different choice of gauges will be emphasized \cite{kallosh}.

\section{KILLING SPINOR ADAPTED GAUGE}

We define the  space-time Killing spinor adapted gauge   as 
follows. First we define an irreducible  $\kappa$-symmetry with only 16 parameters by imposing a condition which $\kappa$ has to satisfy.

\begin{equation}
{\cal P_- } \kappa = {1\over 2} (1- \gamma) \kappa =0
\label{irr}\end{equation}

Now we can use this 16-dimensional local fermionic symmetry to eliminate 16 of $\theta's$.

\begin{equation}
{\cal P_+ } \theta = {1\over 2} (1+ \gamma) \theta =0 \label{g}\end{equation}

The choice of the 
 $\xi^i$-independent projectors
$
{\cal P_{\pm} } = {1\over 2} \Bigl( 1\pm \gamma \Bigr) $
which can divide a 32-dimensional spinor  into 2 parts is made with the help of a Killing spinor of the space-time configuration describing a particular extended object which we are quantizing here. This means that 
\begin{equation}
\gamma=  \Gamma|_{\rm cl} \, ,
\end{equation}
i.e. the constant projector $(1+ \gamma) $ can be chosen as the $\kappa$-symmetry projector  $(1+
\Gamma)$ taken at the values of fields which form a classical solution
describing the relevant bosonic brane. For example, the classical solution for the p-brane is

\begin{equation}
X^i = \xi^i, \qquad X^{a'} = {\rm const} , \qquad \theta=0
\label{cl}\end{equation}
We have split here the bosonic coordinate of the space-time $X^m$ into the part $X^i$ which is identified with the bosonic coordinate of the brane $\xi^i$ and the coordinates $X^{a'}$ which are transverse to the brane  which will become the worldvolume scalars. In the classical solution identifying the Killing spinor of the space-time those are taken to be constants.

The Killing spinor of a space-time
geometry naturally can not depend on the coordinates of the worldvolume.

In a basis in which 
 $\gamma$ is diagonal all  spinors split accordingly:
$$
\theta = \left (\matrix{
\theta ^\alpha \cr
\theta ^{\alpha'} \cr
}\right ) \ ,  \quad \kappa = \left (\matrix{
\kappa ^\alpha \cr
\kappa ^{\alpha'} \cr
}\right ) \ , \quad \epsilon = \left (\matrix{
\epsilon ^\alpha \cr
\epsilon ^{\alpha'} \cr}\right ) 
$$

In this basis the space-time Killing spinor adapted gauge (\ref{irr}), (\ref{g}) takes a simple form
\begin{equation}
 \kappa ^{\alpha'} =0 \ , 
 \qquad \theta ^{\alpha} =0
\end{equation}

 In the same  basis  the field dependent $\kappa$-symmetry generator  $\Gamma$ has the following block structure 
\begin{equation}
 \Gamma = \pmatrix{+C&
(1- C^2) A^{-1}\cr
\cr                A &- C}\, , \end{equation}
where the $16\times 16$ dimensional matrices $C$ and $A$ commute,
$
AC-CA=0
$.
The matrices  $1 \pm\Gamma$ have  vanishing determinants and rank 16, which
means that the 16-dimensional matrices $1\pm C$ and $A$ are invertible.

 If some  choice of a projector ${\cal P_\pm }$ leads to non-invertible A
or
$1\pm C$, this projector   can not be used for quantization. Examples include
d=10 Lorentz covariant gauges for the Green-Schwarz string.

 Consider now a combination of 16-dimensional irreducible $\kappa$-symmetry
and 32-dimensional space-time supersymmetry in our gauge
\begin{eqnarray}\label{keep}
\delta_{\kappa ,  \epsilon} \theta ^\alpha  &=& (1+C)^\alpha{}_\beta
\kappa^\beta + \epsilon ^\alpha =0 \\
\nonumber\\
\delta_{\kappa ,  \epsilon'}  \theta ^{\alpha'} &=& A ^{\alpha'}{}_\beta
\kappa^\beta + \epsilon^{\alpha'}\label{susy}
\end{eqnarray}

One has to impose  the relation between the parameters of local $\kappa$-symmetry  and global space-time supersymmetry $\epsilon$ which will keep the
gauge $\theta ^{\alpha} =0$. It is obtained from eq. (\ref{keep}) under condition that the matrix $(1+C)$ is invertible.
\begin{equation}
 \kappa^\beta (\xi) =- [(1+C)^{-1}] ^\beta{}_ \alpha \epsilon ^\alpha
\label{kappa}\end{equation}
Note that the matrix $C$ is a function of the fields of the theory and therefore
depends on the coordinated of the brane.

\section{WORLDVOLUME SUPERSYMMETRY ON THE BRANE }
 Thus we have explicitly gauge-fixed so far only the fermionic symmetry by requiring $\theta^\alpha =0$. In what follows we will also specify the gauge-fixing of the reparametrization symmetry by choosing the so-called static gauge. This gauge may also be qualified as a Killing vector adapted gauge, as the directions along the branes are in fact related to Killing vectors of the space-time brane configuration. 

\begin{equation}
X^i (\xi ) -  \xi^i =0
\end{equation}
If the reparametrization symmetry is fixed by choosing
a static gauge, the space-time spinors (former scalars on the worldvolume)
like  $\theta ^{\alpha'}$ and
$\epsilon ^\alpha , \epsilon^{\alpha'}
$ become worldvolume spinors.

To extract the 32-dimensional global supersymmetry
transformation of  16  $ \theta ^{\alpha'}$ living on the brane we have to insert the value of the $\kappa$-symmetry parameter transformations (\ref{kappa}) which keeps the $\theta^\alpha =0$ gauge into eq. (\ref{susy})

\begin{equation}
\delta_{\epsilon , \epsilon'}  \theta ^{\alpha'} = \left( - A ^{\alpha'}{}_\beta
[(1+C)^{-1}] ^\beta{}_ \alpha \epsilon ^\alpha\right)_{\rm cl} + \epsilon^{\alpha'}
\label{ans}\end{equation}

This is the {\it general answer} \cite{KalM5}. The subscript $()_{\rm cl}$ means that the values of the matrices $A, C$  have to be taken at the classical values of the brane fields as given e. g. in eqs. (\ref{cl}).

Given a $\kappa$-symmetry of the action is
known and the right choice of the constant projector is made, which supplies us
with $16\times 16$ matrices $C$ and $A$, we have the answer for the worldvolume
supersymmetry. The meaning of the matrices $A$ and $C$ can be understood if we represent the full generator $\Gamma$ as consisting of 2 parts $ \Gamma =  \Gamma_A +  \Gamma _C $ where $   \Gamma_A $ anticommutes with the projector $\gamma$ and $\Gamma _C $ commutes with it.
$$
  \Gamma_A = \pmatrix{0&
(1- C^2) A^{-1}\cr
\cr                A &0}\, , $$
and 
$$
   \Gamma _C =  \pmatrix{+C&
0\cr
\cr                0 &- C} \, . $$
This is in the basis where 
$$
  \gamma = \pmatrix{1&
0\cr
\cr                0 &-1}\, , $$
Thus 
$$
\{ \gamma \ , \Gamma_A \} =0 \ ,  \qquad [\gamma \ , \Gamma_C ] =0 \ .
$$

The condition for the gauge-fixing to be admissible and to produce an action with global supersymmetry is that $A$ and $1\pm C$ are invertible.

\section{EXAMPLES} 

\subsection{GS Superstring}

GS superstring belongs to a class of $\kappa$-symmetric p-brane actions. The $\kappa$-symmetry generator is 
$$
\Gamma = {1\over 2! \sqrt {-g} }\epsilon^{ij}  \gamma_{ij}
$$

Here $\gamma_i = \Pi_i{}^m\Gamma_m$ and $\Pi_i{}^m = \partial _i X^m - \bar \theta \Gamma^m \partial_i \theta$. We will make a  split $m=i, a'$ where $i=0,9$ and $a'=1,\dots , 8$. The Killing spinor 
gauge for the fundamental string is in fact exactly the light-cone gauge for the spinors. Indeed we take as a classical solution
\begin{equation}
X^0 = \tau , \qquad X^9 =\sigma \ , X^{a'} = {\rm const} , \quad \theta=0
\label{clGS}\end{equation}
and the $\kappa$-symmetry generator  at this point in configuration space gives us a projector 
$
1+ \gamma=  1+ \Gamma|_{\rm cl}= 1+ \Gamma_{09} \, ,
$
Thus we have to require that 
\begin{equation}
(1+ \Gamma_{09} ) \theta =0 \qquad (1- \Gamma_{09} ) \kappa=0
\end{equation}
which is equivalent to the standard light-cone gauge for the spinors
\begin{equation}
 \Gamma^+  \theta =0 \qquad  \Gamma^-  \kappa=0
\end{equation}
We may in addition  choose the static gauge for the string $
X^0 = \tau , \qquad X^9 =\sigma \ $. If we start with the version of the theory where the worldsheet metric is the induced one,  we will bring the theory to the form of a 2-dimensional non-linear action of the scalar multiplet which depends on 8 2-dimensional scalars and 8 physical  spinors. The amount of 
unbroken supersymmetries is 32. To find out the supersymmetry transformations of the spinors $\Gamma^-\theta = \theta^{\alpha'}$ we have to find out the relevant $A$  and $C$ matrices.  Clearly, $\Gamma_{ij}, \Gamma_{a'b'}$ commute with $\Gamma_{ij}$ and $\Gamma_{ia'}  $ anticommute. In Killing spinor and vector adapted gauge $\Pi_i{}^j = \delta_i{}^j  - \bar \theta \Gamma^j \partial_i \theta$ and $\Pi_i{}^{a'}  = \partial _i X^{a'}$. This gives us the presentation of the full $\Gamma$ in terms of commuting and anticommuting parts: for $ \Gamma_C$ we find 
$$
  \epsilon^{ij} ( \delta_i{}^{i'}  - \bar \theta \Gamma^{i' }\partial_i \theta) ( \delta_j{}^{j' } - \bar \theta \Gamma^{j' }\partial_j \theta)  \Gamma_{i'j'}+ $$
$$\epsilon^{ij} \partial _i X^{a'} \partial _j X^{b'} \Gamma_{a'b'}
$$
and  $\Gamma_A$ is
$$
  \epsilon^{ij} ( \delta_i{}^{i'}  - \bar \theta \Gamma^{i' }\partial_i \theta) \partial _j X^{b'} \Gamma_{i' b'}$$
Thus the global non-linear worldvolume supersymmetry transformation for the GS superstring quantized in the Killing adapted gauge is given by eq. (\ref{ans}) where the matrices $A,C$ are defined in eqs. above and depend on 
scalars $X^{a'}$ and spinors $\theta^{\alpha'}$. 

\subsection{BST supermembrane}

BST supermembrane also belongs to a class of $\kappa$-symmetric p-brane actions. The $\kappa$-symmetry generator is 
$$
\Gamma = {1\over 3! \sqrt {-g} }\epsilon^{ijk}   \Gamma_{ijk}
$$
To find the Killing spinor adapted gauge for the supermembrane we first consider  the  generator above taken at 
\begin{equation}
X^0 = \tau , \quad X^9 =\sigma \ , \quad X^{10}=\rho \end{equation}
and 
\begin{equation}
 X^{a'} = {\rm const} , \quad \theta=0
\end{equation}
Thus we get 
\begin{equation}
\gamma= \Gamma_{0} \Gamma_{9} \Gamma_{10}
\end{equation}
We see that for the membrane as different from the string the light-cone gauge is not a natural one, as there are 3 Killing directions, time and 2 more and not just one as in case of the string. 
Thus the Killing gauge for the supermebrane is 
\begin{equation}
(1+\Gamma_{0} \Gamma_{9} \Gamma_{10})\theta =0 \ , \quad (1-\Gamma_{0} \Gamma_{9} \Gamma_{10})\kappa =0
\end{equation}
and 
\begin{equation}
X^0 = \tau , \quad X^9 =\sigma \ , \quad X^{10}=\rho \end{equation}
The field remaining in the gauge-fixed action are the scalars
$$X^{a'}(\tau,\sigma, \rho)$$ and the fermions 
$$(1-\Gamma_{0} \Gamma_{9} \Gamma_{10})\theta (\tau,\sigma, \rho).$$

The procedure of extracting the full non-linear 3-dimensional action of the scalar supermultiplet follows as explained above. The action is given by the supermebrane action (in a version of the supermembrane theory with the metric induced on the brane, not independent variable)
at the surface defined by the Killing gauge. For the  global supersymmetry we have to find $A,C$ parts of $\Gamma$ and use eq. (\ref{ans}). 

It is useful to point out here that the complete and detailed quantization of the supermebrane was performed so far only in the light-cone gauge for spinors.  

\subsection{M-5-brane and (0, 2)  tensor supermultiplet in d=6 }

We will give here a brief description of gauge-fixing procedure of the
M-5-brane theory which provides the supersymmetric action for (0, 2)  tensor
supermultiplet in d=6. The  manifestly d=6 general coordinate
invariant M-5-brane action was found by  I. Bandos, K. Lechner, A. Nurmagambetov, P. Pasti, D. Sorokin and M.
Tonin \cite{5b}
\begin{equation}
S_{M-5} \Bigl (X^m (\xi), \theta^\mu(\xi), A_{jk} (\xi) , a (\xi) \Bigr )\, .
\end{equation}
The action depends on $Z^M$ , the metric, induced on the brane, on the tensor field with the self-dual field strength and auxiliary worldvolume
scalar field $a(\xi)$ which  serves to achieve the manifestly d=6 general coordinate invariance of the M-5-brane action.

The $\kappa$-symmetry transformations and supersymmetry of space-time fermions
are
\begin{equation}
\delta_{\kappa, \epsilon } \theta = (1+ \Gamma)\kappa + \epsilon
\end{equation}
where
\begin{equation}
\Gamma=\Gamma_{(0)}  + \Gamma_{(3)}\,  \\
\end{equation}
and
\begin{eqnarray}
\Gamma_{(0)}&=& {\textstyle\frac{1}{6! \sqrt{|g|}}} \epsilon^{i_1\cdots i_6}
\gamma_{i_1}\cdots
\gamma_{i_6}\\ 
 \Gamma_{(3)}&=&{1\over 2\cdot 3!} h_{ijk} \gamma^{ijk}
\nonumber\\
\end{eqnarray}
Here
$i=0,\dots ,5 \ ,  \qquad m=0,\dots , 10 .$

Here we are using the form of
$\kappa$-symmetry transformations found originally in
the superembedding approach by  P.S. Howe, E. Sezgin and P.C. West \cite{HSW} and proved later to be also a symmetry
transformation of the M-5-brane action in \cite{equiv}. The
worldvolume field $h_{ijk}$  of \cite{HSW} turns out to be   a non-linear
function of the fields in the action, whose explicit form can be found in
\cite{equiv}.

The gauge-fixing of the M-5-brane action is inspired by the superembedding
\cite{HSW} of the space-time superspace with coordinates $X^m, \Theta^\mu$ into
worldvolume superspace with coordinates $\xi^i, \theta^\alpha$. We split $m=(i,
a'), \mu = (\alpha, \alpha')$. The superembedding is $X^i = \xi^i \ ,
\Theta^\alpha = \theta^\alpha$ and $X^{a'} =X^{a'}(\xi, \theta) \ ,
\Theta^{\alpha'}= \Theta^{\alpha'}(\xi, \theta)$. To be as close to this as
possible in the bosonic action of the 5-brane we have to require that in our
action
\begin{equation}
X^i = \xi^i, \; \theta^\alpha=0
\label{gauge}\end{equation}
and the fields of the (0,2) tensor multiplet remaining in the action which
depend on $\xi$  are
\begin{equation}
X^{a'}(\xi ) , \; \theta^{\alpha'}(\xi) , A_{ij}(\xi )  , a(\xi ) \ , 
\end{equation}
$$a'=1,2,3,4,5 \ , \qquad \alpha' = 1,2,\dots ,16.$$
Thus we have 5 scalars $X^{a'}(\xi )$, a 16-component spinor $\;
\theta^{\alpha'}(\xi)$ which can considered (see below) as a  a chiral d=6
spinor with a $USp(4)$ symplectic  Majorana-Weyl reality condition
$\theta^{\hat \alpha}_s$, a tensor $A_{ij}(\xi ) $ with the self-dual field
strength and an auxiliary scalar $a(\xi )$.
The 11d  $32\times 32$ $\Gamma^m$ matrices have to be taken in the basis which
correspond to the split of the target superspace into the superspace of the
5-brane and the rest \cite{HSW}.  This reflects the
 $Spin (1,5)\times USp(4)$ symmetry of the six dimensional theory.  An 11d
Majorana spinor decomposes as
$
 \psi = (\psi_{\hat \alpha s} , \; \psi^{\hat \alpha}_s)
$
where $s=1,2,3,4$ is an $USp(4)$ index and $\hat \alpha=1,2,3,4$ is a 6d Weyl
spinor index with upper (lower) indices corresponding to anti-chiral (chiral)
spinors respectively. The 6d spinors satisfy a Majorana-Weyl reality condition.
The relevant representation of 11d   $\Gamma^m$ is
$
\Gamma^i _{\hat \alpha s, \hat \beta t} = \eta_{st} (\sigma^i)_{\hat \alpha
\hat \beta}
$
where $\eta_{st}$ is the $USp(4)$ antisymmetric invariant metric and $\sigma^i$
the 6d chirally-projected gamma-matrices etc. \cite{HSW}. In terms of
16-component spinors we have $ \psi_{\hat \alpha s}=\psi ^{\alpha' }, \;
\psi^{\hat \alpha}_s= \psi^{\alpha}.
$

Thus  we choose a projector $\gamma$ to be  a chiral projector of the
6-dimensional space times the unit matrix. In the basis above this means that
\begin{equation}
\gamma= \Gamma|_{cl} = {\textstyle\frac{1}{6! }} \epsilon^{i_1\cdots i_6}
\gamma_{i_1}\cdots
\gamma_{i_6}|_{cl}  = \pmatrix{
1 &  0 \cr
0 & \ -1 \cr
}\ .
\end{equation}
 Here the subscript ${}_{cl} $ means that we take $X^{a'}_{cl}={\rm const}, \;
\theta_{cl}=0, (h_{ijk})_{cl} =0,  (E_{i} {}^j)_{cl}= \delta _i{}^j,  (E_{ i}
{}^{a'})_{cl}=0 $ and the field-independent part of the $\kappa$-symmetry
generator $1+ \Gamma$ provides us with the projector for gauge-fixing
$\kappa$-symmetry. Note that this is exactly the projector which specifies the
M-5-brane Killing spinor in the target space.

The gauge fixed theory is given by the classical action in the gauge
(\ref{gauge}). Since both the reparametrization symmetry  and $\kappa$-symmetry
are fixed in a unitary way, there are no propagating ghosts. 
The action for a tensor multiplet is an action of a M-5 brane in a Killing
spinor adapted gauge with vanishing $X^i(\xi)-\xi^i$ and $  \theta^\alpha (\xi) $:

$$
S_{(0,2)}  \Bigl (X^{a'}  (\xi), \theta^{\alpha'} (\xi), A_{jk} (\xi) , a (\xi)
\Bigr )=$$
$$S_{M-5} \Bigl (X^{a'} ,  \theta^{\alpha'}  A_{jk}  , a  ,
X^i=\xi^i , \theta^\alpha  =0 \Bigr ) \ .
$$

To find   the exact non-linear worldvolume supersymmetry transformations of the
$S_{(0,2)}$  action
we may now proceed using the rules from the previous section. In addition to
gauge fixing the spinor theta we have to gauge fix the infinite reducible
$\kappa$-symmetry. We choose as before
\begin{equation}
\theta^\alpha=0\ ,  \qquad \kappa^{\alpha'} =0
\end{equation}
To extract from the generator of  $\kappa$-symmetry $\Gamma$ the matrices $C$
and $A$ which define the worldvolume supersymmetry we have to take into account
that  in the flat 11-dimensional background
$$
\Gamma= {\textstyle\frac{1}{6! \sqrt{|g|}}} \epsilon^{i_1\cdots i_6} (
\gamma_{i_1}\cdots
\gamma_{i_6}\   + 40 \gamma_{i_1} \gamma_{i_2}
\gamma_{i_3} h_{i_4 i_5 i_6})  \\
$$
\begin{equation}
 \gamma_i = (\delta_i{}^j - i \bar \theta \Gamma^j \partial_i \theta )
\Gamma_j+ \partial _i X^{a'} \Gamma_{a'}\ .
\label{gamma}\end{equation}
Here we used the fact that the spinors are chiral and therefore $\bar \theta
\Gamma^{a'} \partial_i \theta $ vanishes. Using eq. (\ref{gamma}) we may
rewrite $\Gamma$ as a sum of products of $\Gamma$'s
$$
\Gamma= \sum_{n}
 \Gamma^{i_1} \cdots \Gamma^{i_n} F_{i_i \cdots i_m} (X^{a'}, \theta^{\alpha'},
h_{ijk} )
$$
All terms with even number $n=2,4,6 $ of 6 $\Gamma^{i}$ will contribute only to
$C$,
all terms with odd number $n=1,3,5$  of  $\Gamma^{i}$ will contribute to $A$
since
$\Gamma^{i}$ is off-diagonal in our basis. The dependence on diagonal matrices
$$
\sum_{n=2,4,6}
 \Gamma^{i_1} \cdots \Gamma^{i_n} F_{i_i \cdots i_m} (X^{a'}, \theta^{\alpha'},
h_{ijk}, a )
  \, ,
$$
and $\Gamma_A   $ is 
$$
\sum_{n=1,3,5}
 \Gamma^{i_1} \cdots \Gamma^{i_n} F_{i_i \cdots i_m} (X^{a'}, \theta^{\alpha'},
h_{ijk}, a )
  \, .
$$
Finally  the 32-dimensional supersymmetry transformation  on the brane is given
by the universal eq. (\ref{ans})
with $A(X^{a'}, \theta^{\alpha'}, h_{ijk}, a )
 $ and $C(X^{a'}, \theta^{\alpha'}, h_{ijk}, a )
 $ presented for the M-5-brane above.

The supersymmetry transformations of the bosonic fields, 5 scalars and a
tensor, can be obtained using  the combination of $\kappa$-symmetry and
space-time supersymmetry of these fields and the expression (\ref{susy}).
The linearized form of the worldvolume supersymmetry of the (0,2) tensor
multiplet was given in \cite{HSW}.  In notation appropriate to a 6-dimensional
theory for $\epsilon'=0$
$$
\delta_{\epsilon}  \theta ^s_{\hat \beta} = \epsilon ^{\hat \alpha t} \Bigl
({1\over 2}
\sigma^i _{\hat \alpha \hat \beta} (\gamma_{b'})_t{}^s \partial_i X^{b'} -
{1\over 6} (\sigma^{ijk}_{\hat \alpha \hat \beta} \delta_t{}^s h_{ijk}\Bigr) \ .
$$
One can recognize here terms linear and cubic in $\Gamma^i$ which form the
linear approximation of our matrix $A$.

Note however that the full non-linear action of the self-interacting tensor
multiplet has also a symmetry under additional 16-component chiral spinor
$\epsilon^{\alpha'}= \epsilon_{\hat \alpha s}$.  The one with the anti-chiral
spinor
$\epsilon^{\alpha}= \epsilon^{\hat \alpha}_s $   in the linear approximation
relates the spinor of the tensor multiplet to the derivative of  scalars and to
the tensor field strength. The non-linear action has both chiral as well as
anti-chiral supersymmetries.

\section{GENERAL CLASS OF GAUGES }

In general, there is no need to use only the class of gauges defined above which are related to Killing spinors. A much larger class of gauges is available for the quantization of any of the p-branes, D-p-branes and M-branes.  This amounts to say that the choice of a projector $\gamma$ can be arbitrary as well as the choice of the gauge which fixes the reparametrization symmetry. In particular, in case of D-p-branes the advantage can be taken from the fact that the quantization consistent with Lorentz symmetry is available \cite{Ag1,bk,kallosh}. In some cases the advantage is to bring the quantized action to a free  quadratic action, as in the case of the D-0-brane. This requires a special choice of the gauge for reparametrization symmetry. In generic situation we may formulate here the basic constraint on the choice of the gauge to fix the $\kappa$-symmetry. 

Assume that we have chosen the projector $1\pm \gamma$  and fixed the reparametrization symmetry arbitrarily. 

The condition of consistency of this choice can be formulated in a basis  where  $\gamma$ has 16 of $+1$ and 16 of $-1$ on the diagonal.

 In this  basis  the field dependent $\kappa$-symmetry generator  $1\pm \Gamma$ has the following block structure 
\begin{equation}
1\pm \Gamma = \pmatrix{1\pm C&
(1- C^2) A^{-1}\cr
\cr                A &1\mp C}\, , \end{equation}
The field dependent matrices  $1 \pm\Gamma$ have  vanishing determinants and rank 16. For this to happen,   the 16-dimensional matrices $1\pm C$ and $A$ have to be invertible. This has to be true in any gauge which is chosen and not only in the Killing gauge, as described above.

For some  choice of a projector ${\cal P_\pm }$ with particular combination of the reparametrization gauge one may encounter a 
 non-invertible $A$ and/or
$1\pm C$.  This projector ${\cal P_\pm }$ together with this choice of the reparametrization gauge can not be used for quantization. 

The global supersymmetry on the worldvolume of the 16-component spinor  $\theta$ is given by 

\begin{equation}
\delta  \theta  = (- A 
[(1+C)^{-1}])_{g.f.}    \epsilon + \epsilon'
\label{ansGen}\end{equation}

Here the subscript ${}_{g.f.}$ means that the field dependent $A,C$ matrices have to be taken at the values of the fields with some particular gauge-fixing of reparametrization symmetry.

For example for D-p-branes  for $p$ even there is a choice $\gamma= \Gamma_{11}$,  $C=0$
and an invertible $A$ can be find in \cite{Ag1,bk,kallosh} together with the
total procedure of quantization. This is an example when both
$1\pm C$ and $A$ are invertible for a given choice of $\gamma$.
With the same choice of $\gamma$, which gives the  only possible Lorentz covariant splitting of the 32-dimensional spinor in half in d=10,   type IIA GS string will have a non-invertible
$A=0$ and this gauge is not acceptable as it is known for a very long time.
This was always a major obstacle for manifestly supersymmetric Lorentz covariant superstring theory quantization.

\subsection{Covariant Quantization of the D-0-brane}

Consider the $\kappa$-symmetric action of a D-0-brane. D-0-brane action does
not
have Born-Infeld field since there is no place for an antisymmetric tensor of
rank 2 in one-dimensional theory.
 The action for $p=0$ case reduces to
\begin{equation}
S = -T \left (  \int d\tau \sqrt { - G_{\tau\tau}}
  +  \int  \bar \theta \Gamma^{11} \dot \theta \right) \ .
\label{0action}\end{equation}
This action can be derived from the action of the massless 11-dimensional
superparticle.
\begin{equation}
S = \int d\tau \sqrt {g_{\tau\tau}}  g^{\tau\tau} \left( \dot X^{\hat m}  -
\bar \theta \Gamma^{\hat m} \dot \theta \right)^2 \ , \label{11}\end{equation}

Here $ \hat m =
 0,1,\cdots , 8,9,10.
$
We may solve  equation of motion for $X^{ \hat {10}}$    as ${\bf P}_{ \hat
{10}} =Z$, where $Z$ is a constant, and use   $ \Gamma^{11} =
\Gamma^{\hat {10}}
$. From this one can deduce a
 first order action
\begin{equation}\label{first}
S = \int d\tau ( {\bf P}_{m} (\dot X^m - \bar \theta \Gamma^m \dot \theta)
\end{equation}
$$ +{1\over 2} V (
{\bf P}^2 + Z^2) - Z \bar \theta \Gamma^{11} \dot \theta + \bar \chi_1
d_2) \ .
$$

The action  (\ref{first}) is invariant under the 16-dimensional irreducible
$\kappa$-symmetry and under the
reparametrization symmetry. The gauge symmetries are (we denote $\Gamma^m {\bf
P}_{m}= / {\hskip - 0.27 cm  {\bf P}}$):
$$
\delta \bar \theta= \bar  \kappa_2 ( \Gamma^{11} Z + / {\hskip - 0.27 cm  {\bf P}}) \ , $$
$$
\delta X^m = -\eta {\bf P}^{m}  - \delta \bar \theta  \Gamma^m \theta - \bar
\kappa_2 \Gamma^m d \ ,$$
$$
\delta V = \dot \eta + 4 \bar \kappa_2 \dot \theta + 2 \bar \chi_1 \kappa_2\
, $$
$$
\delta \bar \chi = \bar \kappa_2  \;  \dot / {\hskip - 0.27 cm  {\bf P}} \ ,$$
$$
\delta d =[ {\bf P}^2 + Z^2] \kappa_2 \ .
$$
Here $\eta(\tau) $ is the time reparametrization gauge parameter and
$\kappa_2(\tau)
=
{1\over 2} (1- \Gamma^{11} ) \kappa (\tau) $ is the 16-dimensional parameter of
$\kappa$-symmetry. The gauge symmetries form a closed algebra
$
[\delta({\kappa_2}) ,\delta({\kappa'_2}) ] = \delta ({\eta = 2 \bar \kappa_2
\sl \kappa'_2} ) \ .
$

To bring the theory to the canonical form we introduce  canonical momenta to
$\theta$ and to $V$ and find, excluding auxiliary fields
\begin{equation}
L =  {\bf P}_{m} \dot X^m+ P_V \dot V + \bar P_{\theta } \dot \theta  
\end{equation}
$$+ {1\over
2} V ({\bf P}^2
 + Z^2)
  +   P_V \varphi + \left (\bar P_{\theta }  + \bar \theta  (/ {\hskip - 0.27 cm  {\bf P}} + Z
\Gamma^{11})\right ) \psi \ .$$

We have  primary constraints $\bar \Phi  \equiv  \bar P_{\theta }  + \bar
\theta  (/ {\hskip - 0.27 cm  {\bf P}} + Z \Gamma^{11}) \approx 0$
and $P_V=0$. The Poisson brackets for 32 fermionic constraints are
$
\{ \Phi  ,  \Phi  \}=  2 C ( / {\hskip - 0.27 cm  {\bf P}}  + \Gamma_{11} Z ) \ .
$
We also have to require that the constraints are consistent with the time
evolution $\{P_V, H\}=0$. This generates a secondary constraint
\begin{equation}
t=  {\bf P}^2 + Z^2 \ .
\end{equation}
Thus the Hamiltonian is weakly zero and any physical state of the system
satisfying the reparamet-\\
rization constraint is a BPS state $M=|Z|$ since
\begin{equation}
 {\bf P}^2 + Z^2 | \Psi \rangle  = 0 \qquad  \Longrightarrow  \qquad 
\end{equation}
$$Z^2   |
\Psi \rangle =-  {\bf P}^2  | \Psi \rangle = M^2  |  \Psi  \rangle \ .
$$
 The $32\times 32$ -dimensional matrix $C ( / {\hskip - 0.27 cm  {\bf P}} + \Gamma_{11} Z) $ is not
invertible since it squares to zero when the reparametrization constraint is
imposed. This is a reminder of the fact that D-0-brane is a d=11 massless
superparticle.
The 32 dimensional fermionic constraint has a 16-dimensional part which forms a
first class constraint and another 16-dimensional part which forms a second
class constraint.
We notice that the Poisson brackets reproduce the $d=10$, $N=2$ algebra with
the
central charge which can also be understood as $d=11$, $N=1$ supersymmetry
algebra with the
constant value of ${\bf P}_{11}=Z$.

We proceed with the quantization and gauge-fix
$\kappa$-symmetry covariantly   by taking $\theta_2=0, \theta_1 \equiv \lambda
$
and find
\begin{equation}
L^\kappa_{g.f.}= {\bf P}_{m} (\dot X^m - \bar \lambda  \Gamma^m \dot \lambda )
+ {1\over 2} V (
{\bf P}^2 + Z^2) \ .
\end{equation}
The 16-dimensional fermionic constraint
\begin{equation}
 \bar \Phi_{\lambda }  \equiv  (\bar P_{\lambda }  + \bar \lambda  / {\hskip - 0.27 cm  {\bf P}} )\approx
0
\end{equation}
forms the Poisson bracket
\begin{equation}
\{  \Phi^{\alpha }_{\lambda }  ,  \Phi ^{\beta }_{\lambda } \}=   2  ( / {\hskip - 0.27 cm  {\bf P}}
C)^{\alpha \beta}   \label{Pois} \ .
 \end{equation}
The matrix $ / {\hskip - 0.27 cm  {\bf P}} C
$ is perfectly invertible as long as the central charge $Z$ is not vanishing.
The inverse to (\ref{Pois}) is
\begin{equation}
\{  \Phi^{\alpha }  ,  \Phi ^{\beta } \}^{-1}  \mid_{t=0} \;
= [2  (  / {\hskip - 0.27 cm  {\bf P}} C)^{\alpha \beta}   ]^{-1} = {(C  / {\hskip - 0.27 cm  {\bf P}} )_{\alpha \beta}
 \over 2  {\bf P}^2} \ .
  \end{equation}

This proves that the fermionic constraints are second class and that the
fermionic part of the Lagrangian
is not degenerate in a Lorentz covariant gauge. None of this would be true for
a vanishing central charge. Note that in the rest frame ${\bf P}_{0}=M , \vec
{\bf P}=0
$, hence
\begin{equation}
 \Phi_{\alpha \beta} = M \delta_{\alpha \beta} \ .
\end{equation}
For D-0-brane one can covariantly gauge-fix the reparametrization symmetry by
choosing the $V=1$ gauge and including the anticommuting reparametrization
ghosts
$b,c$. This brings us to the following form of the action:
\begin{equation}
L^{\kappa, \eta} _{g.f.}= {\bf P}_{m} \dot X^m - \bar \lambda  / {\hskip - 0.27 cm  {\bf P}} \dot \lambda
 + {1\over 2}  (
{\bf P}^2 + Z^2)
 +  b\dot c \ .
\end{equation}
Now we can define Dirac brackets
\begin{eqnarray}
 \{  \lambda  ,  \bar \lambda \}^*= \{  \lambda , \bar \Phi \}  \{ \bar \Phi_
,  \Phi  \}^{-1}  \{  \Phi  \ , \bar \lambda \} 
\end{eqnarray}$$
= {  / {\hskip - 0.27 cm  {\bf P}} \over 2 {\bf P}^2} =-
{ / {\hskip - 0.27 cm  {\bf P}} \over 2 Z^2} \ .
$$
The generator of the 32-dimensional supersymmetry is
$
\bar \epsilon Q = \bar \epsilon ( / {\hskip - 0.27 cm  {\bf P}} + \Gamma^{11} Z) \lambda \ .
$
It forms the following Dirac bracket
\begin{equation}
[\bar \epsilon Q \ , \bar Q \epsilon' ]^* = \bar \epsilon ( / {\hskip - 0.27 cm  {\bf P}} +
\Gamma^{11} Z)  {    / {\hskip - 0.27 cm  {\bf P}} \over 2 {\bf P}^2} ( / {\hskip - 0.27 cm  {\bf P}} +
\Gamma^{11} Z)  \epsilon'=
\end{equation}
$$
\bar \epsilon \Gamma^{\hat m}  {\bf P}_{\hat m}   \epsilon'   = \bar \epsilon
 / {\hskip - 0.27 cm  {\bf P}} +
\Gamma^{11} Z)  \epsilon' \ .
\label{diracsusy}$$

This Dirac bracket realizes the d=11, N=1 supersymmetry algebra
 or, equivalently, d=10, N=2 supersymmetry algebra with the central charge
$Z$.

One can also  to take into account that the path integral in presence of second
class constraints has an additional term with $\sqrt{{\rm Ber} \{\Phi _\lambda
,
\Phi_\lambda  \}} \sim  \sqrt{ {\rm Ber}\, \Phi_{\alpha \beta} }$~
as it is known from the work of E. Fradkin and collaborators. It can be
used to make a change of variables
$
S_\alpha  = \Phi^{1/2}_{\alpha \beta}    \; \lambda^\beta \ .
$
The action becomes
\begin{equation}
L= {\bf P}_{m} \dot X^m-  i S_\alpha  \dot S_\alpha   +  b\dot c - H
\end{equation}
\begin{equation}
H= - {1\over 2} (
{\bf P}^2 + Z^2) \ .
\end{equation}
The generators of global supersymmetry commuting with the Hamiltonian take the
form
\begin{equation}
\bar  \epsilon  Q    =   \bar \epsilon (/ {\hskip - 0.27 cm  {\bf P}} + \Gamma^{11} Z)  \Phi^{-1/2} S \
{}.
\end{equation}
Taking into account that $ \{ S _\alpha  ,  S_\beta  \}^* =-{i\over 2}
\delta_{\alpha \beta}$
we have again realized $d=10$, $N=2$  supersymmetry algebra in the form
(\ref{diracsusy}) or  (\ref{diracsusy11}).

The terms with anticommuting fields $S_\alpha $ can be rewritten in a form
where it is clear that they can be interpreted as   world-line spinors,
 \begin{equation}
L= {\bf P}_{m}  \partial_0 X^m +  \bar S_\alpha  \rho^0 \partial_0  S_\alpha
+  b\dot c - H \ .
\end{equation}
Here $\bar S_\alpha =  i S_\alpha  \rho^0$ and $(\rho^0)^2 =-1$, $\rho^0=i$
being a 1-dimensional matrix.

Thus, we have the original 10 coordinates $X^m$ and their conjugate momenta
${\bf P}_{m}$, and a pair of reparametrization ghosts. There are also 16
anticommuting world-line spinors $S$, describing 8 fermionic degrees of
freedom. The Hamiltonian  is quadratic.
The  ground state with   $$
M^2=Z^2$$
 is the state with the minimal value of the
Hamiltonian.
Thus for the D-superparticle one can see that the condition for the covariant
quantization is satisfied in the presence of a central charge which makes the
mass of a physical state non-vanishing \cite{kallosh}. The global supersymmetry algebra is
realized in a covariant way, as different from the light-cone gauge.

Note that the total procedure of quantization has new features due to the fact that we deal with BPS states with the non-vanishing mass. In particular we have divided on $Z^2$ at the intermediate level, which would be impossible at $Z=0$. The BPS states with vanishing mass have to be considered separately and with  great care, as we know it from the analysis of the massless black holes in string theory \cite{strom}.

 \section {SUMMARY OF NEW RESULTS}

In this talk we have described various new possibilities of quantization of the fermionic worldvolume gauge symmetry, so called $\kappa$-symmetry. Our main focus was on relation between the space-time supersymmetric configurations, or BPS configurations and the new ways of quantization. 

It was known for a long time that the superparticle and the superstring with manifest space-time supersymmetry can not be quantized covariantly. Now we have learned that both the D-0-brane and D-1-string as well as all of the D-p-branes do admit covariant gauge-fixing. 

We have also used our experience in the properties of the Killing spinors of various BPS solutions in the space-time. We have applied this concept to the 
quantization of the Green-Schwarz Superstring and Bergshoeff-Sezgin-Townsend Supermembrane. This results in an interesting and  yet unexplored non-linear theories of the scalar supermultiplets in d=2 and d=3 dimensions. 

One of the most interesting new results is related to the quantization of the M-5-brane. The quantization is most fruitful in the fermionic gauge defined by the Killing spinor  and reparametrization gauge defined by the Killing vector of the space-time M-5-brane solution. The resulting action defines the non-linear action of the (2,0) d=6 tensor supermultiplet.

\end{document}